# The Identification of Biological Stains at Crime Scenes: A Promising Role for Proteomics and Machine Learning

Anna Rosenberg[†‡], Stéphanie Laurent[¶†], Esther Morandeau[¶†], Alix Munoz[†], and Joelle Vinh[*†]

[†]Spectrométrie de Masse Biologique et Protéomique, ESPCI-PSL, CNRS, UAR 2051, Rue Vauquelin, 75005, Paris, France

[‡]Department of biology, Ecole Normale Supérieure Paris-Saclay, Avenue des Sciences, 91190, Gif-sur-Yvette, France

[¶]Département Environnement Incendies Explosifs, Institut de Recherche Criminelle de la Gendarmerie Nationale, Boulevard de l'Hautil, 95000, Cergy-Pontoise, France

**ABSTRACT:** Forensic body fluid identification is crucial for reconstructing crime scene events. While DNA analysis provides individualization, it lacks information about the fluid's origin. We developed and evaluated three complementary proteomic approaches using LC-HRMS/MS to identify blood, saliva, semen, urine, and vaginal fluid, including complex mixtures. The first method utilized fluid-specific peptide biomarkers, achieving high accuracy for pure fluids. The second employed peptide abundance ratios, demonstrating effectiveness in body fluid mixtures. The third, a machine learning model using Classifier Chain Random Forest, achieved 100% accuracy for pure fluids and promising results for mixtures. Our results revealed the complementarity of different tests, with the peptide-specific biomarker and machine-learning approaches being the most robust. This study demonstrates the potential of proteomics for comprehensive body fluid identification, offering valuable tools for forensic investigations.

Even though DNA sequencing is a valuable tool in forensics, it does not provide information about the origin and nature of a stain, which is crucial for reconstructing events. Indeed, in cases of sexual assault involving related individuals, determining whether DNA on the victim's clothing originates from semen or epidermal cells is critical. This necessity has spurred the development of various methods for identifying body fluids.[1] These methods, however, fall into two categories: presumptive and confirmatory.

Presumptive methods include microscopy-based techniques (like spermatozoa identification for semen analysis) and chemical-based approaches (such as Luminol for blood detection).[2,3] They also include biochemical methodologies like immunochromatographic tests, which use specific antibodies to detect proteins in various body fluids (e.g., hemoglobin, α-amylase, prostate-specific antigen, semenogelin I and II, and uromodulin) and are analogous to ELISA tests.[4–12] A recent review evaluating their performance, such as semen and blood tests, was published by Medina-Paz et al. (2024)[13] While these methods are quick and simple, they only provide a preliminary indication of a biological fluid's presence and can sometimes lead to false positives. These tests rely on protein detection, but this process can be compromised by **epitope degradation**, which affects the binding of antibodies to antigens. Furthermore, some tests target non-specific markers. For example, using **non-specific enzymatic activity** from amylase to detect saliva can lead to false positives. Similarly, detecting **prostate-specific antigen (PSA)** for semen can yield misleading results, as PSA can also be present in other body fluids as reported by Yang et al. (2013).[14] They also often target only one or two proteins and are limited to a single body fluid, posing challenges in real-case scenarios.

**Confirmatory methods**, on the other hand, provide definitive results. More recently, RNA-based tests (including messenger RNA and miRNA) and methods based on methylation levels of specific loci using bisulfite sequencing have emerged to address the need for comprehensive fluid identification.[15–18] However, these approaches rely on nucleic acid amplification, which can be affected by environmental conditions, limiting their reliability.[19] In addition, even though these approaches target more than one RNA or CpG site, unlike immunological tests, which target just one protein, they are limited by the necessity of using specific primers and do not provide a comprehensive view of the epigenome or transcriptome. Unlike targeted molecular methods, proteomic mass spectrometry provides a broader, more comprehensive overview, as illustrated in this paper. It offers a significant advantage by enabling conclusive identification of body fluids, even in complex mixtures. This is particularly relevant in the context of forensic casework, where certain body fluids remain challenging to identify. Notably, presumptive tests for vaginal fluid identification remain underdeveloped, despite their potential relevance in sexual assault cases, which further highlights the need for a robust confirmatory approach.[1]

Liquid Chromatography coupled with tandem Mass Spectrometry (LC-MS/MS) enables the exploration of protein composition, offering a quasi-comprehensive proteome profiling, that allows differentiation of body fluids.[20,21] Some proteins are known for their stability and could be used as robust biomarkers.[22] Their level can be highly modulated according to their origin, and some are considered specific to certain fluids.[21] Using bottom-up proteomics enables the characterization of very complex protein mixtures with minimal material (< µg). In recent years, proteomic mass spectrometry has established itself as a valuable complementary forensic tool to DNA profiling, allowing for the identification of not only the presence of biological fluids, but also their origin. Yang et al. (2013) were among the first to demonstrate that a MALDI-MS approach based on protein markers could simultaneously identify multiple matrices (blood, saliva, semen, vaginal secretions, etc.), even in mixtures, while preserving the DNA for later analysis.[14] For their part, Legg et al. isolated a panel of biomarkers highly specific to six types of fluids via a comparative strategy and 2D-HPLC fractionation, paving the way for a high-resolution multiplex approach.[23] Additionally, studies by Illiano et al. (2018) have illustrated the ability of LC–MS/MS in MRM mode to generate targeted peptide signatures even in mixtures,[24] while Kennedy et al. (2020) developed a rapid MALDI-MS method capable of distinguishing between human and animal samples.[25] These advances represent significant progress; however, these methods have been mainly developed and validated on pure biological fluids. Wang et al. (2025) refined these approaches by combining bottom-up proteomics discovery and targeted modes, revealing unique protein markers for peripheral blood, menstrual blood, saliva, semen, and vaginal secretions.[26]

In reality, the major analytical challenge in a forensic context concerns the analysis of biological mixtures deposited on complex substrates—tissues, textiles, or various surfaces—where matrix effects, ionization suppression, spectral complexity, and difficulty in interpreting the data seriously compromise the reliable detection of protein biomarkers.

Multiple data processing strategies can be employed for body fluid identification. One common approach involves identifying specific proteins or peptides that serve as biomarkers—either unique to a given fluid or significantly over- or under-expressed compared to others. Another strategy, commonly used in other fields, consists of applying artificial intelligence: by training machine learning (ML) classifiers on proteomics data, this approach leverages a broader range of signals rather than relying on the expression levels of just a few proteins.[27,28] We present a new method for body fluid identification in this context. We compare three tests on a cohort of 20 individuals: the first relies on peptides specific to one of the five body fluids among blood, saliva, semen, urine, and vaginal fluids, the second assesses peptide relative abundance ratios, and the last uses an artificial intelligence model.

The combination of AI/ML with epigenetic and transcriptomic analysis has recently led to significant advancements in forensic science. AI/ML models enable us to extract information from data for precise and comprehensive analysis, particularly in complex mixtures. Applying metagenomic approaches, Kim et al. (2025) used a Naive Bayes algorithm to analyze the bacterial profiles of various body fluids in a forensic context.[29] The model by Lynch et al. (2024) combines ML algorithms to identify body fluids from mRNA profiling.[30] The multi-class Random Forest (RF) model by Iacob et al. (2019) predicts their origin even in mixtures.[31] The multi-class Support Vector Machine (SVM) model by Li et al. (2025) classifies several body fluids using only 14 specific microRNAs, which shows that a targeted approach using a small panel of microRNAs can provide robust and accurate identification.[32] In the field of epigenetics, Zhao et al. (2025) employ AI-based feature selection methods to identify the most relevant CpG sites for fluid identification.[33]

This work confirms that AI and ML are powerful tools for improving forensic techniques. Besides traditional statistical analysis, AI/ML models provide an alternative that can handle the complexity of biological mixtures and identify unique molecular signatures. We hypothesize that, as a significant step forward in forensic biofluid identification, they will likely become a standard part of forensic practice in the coming years.

## EXPERIMENTAL SECTION

### Materials

Ultrapure solvent and reagents were used: dithiothreitol (DTT, Thermo Scientific, electrophoresis grade 99%), Iodoacetamide (IAM, Thermo Scientific, 98%), ethanol (EtOH, Fisher, HPLC grade 99.8%), ammonium bicarbonate (ABC Sigma-Aldrich, 99%), trypsin + LysC mix (Promega, MS grade), acetonitrile (ACN, Optima LC-MS ThermoFisher), H2O is MilliQ water (Merck Millipore), formic acid (FA, Thermo Scientific, 99%)

### Sample collection

Samples from volunteers were home collected for blood, saliva, semen, urine, and vaginal fluids on sterile swabs (4N6FLOQSwabs® Genetics 4508C) according to the supplier protocol. Blood was collected using a finger prick, with a few drops applied to the swab. Semen and urine were collected by applying them directly onto the swab. The vaginal swab was taken by gently rubbing the swab against the vaginal walls. Saliva was collected by spitting into a tube and then dipping the swab into the saliva. Urine and vaginal fluids were collected outside the menstrual period. Vaginal fluid was sampled after at least 5 days of sexual abstinence. The samples were stored at room temperature by the participants for up to 2 days after collection and then at 4°C until analysis. Written consent was collected from all anonymous participants as mentioned in the ethical recommendations validated by the PJGN (Pole Judiciaire de la Gendarmerie Nationale) ethical committee. Twenty participants were used per fluid, with an equal distribution according to gender for non-sexual fluids. For mixtures, three different individuals per fluid were used. The sex distribution of participants is available in Supplementary Figure S1.

### Protein extraction and quantitation

Samples were extracted using the Single-pot, solid-phase-enhanced sample preparation for proteomics experiment (Sp3) protocol.[34] Swabs were placed into a 1% deoxycholate solution (w/w), and nucleic acids were eliminated by sonication (15 min, VNR #USC200T sonicator bath). For reference mixtures, the fluids were mixed in equal volumes after this step. The 26 sample preparations (*i.e.,* mixtures containing 2, 3, 4, or 5 fluids) were prepared 3 times using fluids from 3 different donors (78 mixtures).

The proteins were reduced and alkylated in ABC 500mM using 5mM DTT and 25mM IAM, extracted on 25 µL of beads (1:1, Cytiva Sera-Mag™ SpeedBead Carboxylate-Modified [E3] and [E7] Magnetic Particles) in EtOH and washed in H2O/EtOH 20/80 (v/v). After overnight proteolysis with a trypsin/LysC mix (10ng/µL in 100 µL ABC 50mM) and removal of the magnetic beads, the proteolytic peptides were dried in a SpeedVac, resuspended in 40 µL of solvent A (0.1% FA in H2O/ACN 98/2 (v/v)), and quantified using NanoDrop™ One (Thermo Scientific, e205=31 method) to adjust sample concentration to 50 ng/µL in solvent A.

### LC-MS/MS analysis

The digests were injected (2µL, 100 ng) in a nanoRSLC U3000 system coupled to a nano-ESI QqOrbitrap hybrid mass spectrometer Q Exactive (ThermoFisher Scientific). Peptides were loaded onto a C18 trap column (Acclaim PepMap100, 300 µm i.d. X 5 mm, 5 µm,100Å, ThermoFisher Scientific), desalted (30 µL.min$^{-1}$, solvent A, 3 min), transferred to a C18 column (nanoEase™M/Z HSS C18 T3, 75 µm i.d. X 250 mm, 1.8 µm, 100Å, Waters) and eluted at a flow rate 220 nL.min$^{-1}$ with a 2–40% B in 60 min gradient of solvent B (0.1% FA in H2O/ACN 10:90 (v/v)). Top 10

DDA mode MS consisted of one MS full scan (m/z 200–2000, resolution 70,000 at 200 m/z, max. injection time 100 ms, AGC 3e6) followed by ten MS/MS acquisitions (isolation windows 4 m/z, resolution 17,500 at 200 m/z, max. injection time 50 ms, AGC 1e5, norm. collision energy 30 and dynamic exclusion of 10 s). The data have been deposited to the ProteomeXchange Consortium repository via the PRIDE partner with the dataset identifier PXD053891.[35]

## COMPUTATION METHODS AND PROGRAMS

### Data analysis

Raw data were processed with Protein Discoverer 2.4.0 (Thermo Scientific), using Sequest™HT (Thermo Scientific) against the Human proteome database (SwissProt v.25/03/2024) with a mass tolerance at 10 ppm for MS and 0.02 Da for MS/MS data, using trypsin digestion with up to 2 miscleavages, Met oxidation and N-terminal acetylation as variable modifications, Cys carbamidomethylation as fixed modification. Peptide Spectrum Matches (PSMs) were filtered using a False Discovery Rate (FDR) of 0.01, calculated with a concatenated target-decoy database.

### Construction of a list of peptides specific to each fluid

Ten individuals for each fluid (5 men and 5 women for non-sexual fluid) were used to compile a list of fluid-specific peptides. Peptide intensity was normalized by the sum of the intensities of all peptides detected in the run. The criteria for peptide selection included 80% sensitivity (detection in 8 out of 10 individuals fluid with normalized intensity > 5E-5) and 90% specificity (detection in no more than 1 out of 10 individuals for each other fluid), see Supplementary Table S1 for candidate peptide list.

The list was further tested with 10 distinct individuals per fluid (50 samples), 3 samples from different individuals for each potential mixture (26 possible mixtures, 78 tested mixtures), and 15 negative controls without any body fluids. The test's performance was assessed by counting the number of expected peptides experimentally detected. We developed a script to analyze sample composition: (1) A threshold for identifying each fluid was determined from the associated receiver operating characteristic (ROC) curve and set to maximize sensitivity and specificity by choosing values closest to the top-left corner of the ROC curve; (2) A fluid is considered identified if the ratio of the number of detected peptides to the total number of expected peptides for that fluid exceeds its respective threshold.

### Construction of a quantitative model for fluid recognition

Ten individuals for each fluid (5 men and 5 women for non-sexual fluids) were used to construct a list of fluid-specific ratios. A peptide is not considered if not identified by MS/MS (abundance is set to 0 in such a case).

Peptides detected in all 50 samples were selected as potential denominators, and peptides detected in at least one sample were selected as potential numerators. From these two lists of peptides, ratios between all pairs of peptides were calculated for each sample. An ANOVA test was conducted for each ratio to compare one fluid against the others. The 50 ratios with the lowest p-value were selected for each fluid (see Supplementary Table S2). A confidence interval of 99,99% was calculated for the other fluids and the associated fluid for each ratio.

The relevance of the ratios was further tested with 10 distinct individuals per fluid (50 samples), 3 samples from different individuals for each potential mixture (26 possible mixtures, 78 tested mixtures), and 15 negative controls without any body fluids.

Test efficiency is evaluated by calculating all ratios and comparing them to the mean for all fluids, according to the confidence interval. For each ratio, if the mean is higher, respectively lower than that of the other ratios, the ratio is counted for the associated body fluid if it falls within the upper, respectively lower, confidence interval limits for both the associated body fluid and the other fluids. We developed a script to test sample composition: (1) A threshold for identifying each fluid was determined by considering the ratio between the experimental number of validated ratios and the total number of calculated ratios (50) for a given fluid by plotting a ROC curve and selecting the optimal sensitivity and specificity values (top-left corner of the ROC curve); (2) A fluid is identified if the ratio of counted ratios over the total ratios for that fluid exceeds its respective threshold.

## Construction of a supervised learning model for fluid recognition

### Datasets used

The complete dataset consisted of samples from 20 individuals for pure fluids, 15 negative controls, and 3 samples for each possible mixture. This dataset was divided into three different sets: a training set with 12 individuals for each fluid, 9 negative controls, and 1 sample for each mixture; a validation set for parameters optimization with 4 individuals for each fluid, 3 negative controls, and 1 sample for each mixture; and an independent test set following the same distribution as the validation set. The distribution details are available in the Excel file on the GitHub repository at https://github.com/SMBP-lab/Body-fluid-identification-proteomics.

### Features selection

Features used were the intensity of peptide signal as estimated by proteome discoverer analysis. We reduced the number of features in the model using two selection methods. Firstly, an ANOVA test was conducted on pure fluids from the training set, and features were sorted by p-value (from lowest to highest).[28] Secondly, a ReliefF-based Multilabel Feature Selection was performed on the training set, and features were sorted by ReliefF score (from highest to lowest).[36–38]

An optimization protocol (as mentioned below) has been realized for the x first features for each selection mode (with x belonging to 100-10,000 with a step of 100) to evaluate the influence of the number of features selected on models used. Hamming loss and subset accuracy of the best model on validation data after optimization during 5 cycles were compared between selection mode and number of features selected.

Based on the initial selection step results, a secondary investigation into the optimal number of features was conducted specifically on ReliefF-selected features to accurately evaluate the impact of the number of features, selecting between 500 and 4,000 features with a step of 20 and 30 optimization cycles.

### Models used for supervised machine learning

This study compared five supervised classifiers: Random Forest (RF), Support Vector Machines (SVM), K-nearest-neighbors (KNN), Logistic Regression (LR), and Gaussian Naive Bayes (GNB)[39–43] Additionally, two different methods for problem transformation in multilabel scenarios were tested: Classifier Chains and Binary relevance.[44–47] The models used were obtained from the scikit-learn and scikit-multilearn libraries of Python.[48,49] Hamming loss and subset accuracy were compared for the best-optimized model (30 optimization cycles) without feature selection for each possibility with one classifier and one model transformation strategy.

### Optimization of parameters

The classifier parameters were optimized using the Optuna framework.[50] For RF, using between 100 and 1000 estimators, the maximum depth ranged from 10 to 50 in increments of 10 or was set to None. The minimum number of samples required to split a node ranged from 2 to 20, and required to be at a leaf node ranged from 1 to 20.

For SVM, the regularization coefficient was tuned from 0.1 to 40. The kernel function was tested with various options, including linear, polynomial (degrees 2 to 5), sigmoid, and radial basis function (RBF). The parameter $\gamma$ for RBF, linear, and sigmoid kernels was defined as 1 divided by either (i) the number of features multiplied by the mean-variance of features (scaled in scikit learn) or (ii) the number of features (auto in scikit learn).

The number of neighbors for the KNN was tuned from 1 to 20. The weight function used in the prediction was optimized between uniform weights or the inverse of each neighbor's distance.

For LR, the C parameter was tuned from 1E-4 to 100 with either Lasso or Ridge regularization.

For the GNB, the smoothing variance was varied from 1E-10 to 1E-2.

There were 30 cycles of optimization for model selection, 5 cycles for step 1 of feature selection, 30

cycles for step 2 (with reliefF-like algorithm selected features and between 500 and 4,000 features), and 50 cycles in final training. Optimization was made by minimizing Hamming loss on the validation set.

The importance of features was calculated by adding the mean decrease in impurity for each of the five Random Forest classifiers in the Classifier Chain.[51,52]

### Evaluation of models

The performance of the models was evaluated using different metrics calculated for all samples, whether single or mixed fluids, using the sample composition. We calculated accuracy, F1-score, precision, and recall for each fluid. Then, micro, macro, and weighted F1-score, precision, and recall were computed. Finally, subset accuracy and Hamming loss were calculated and compared between models. All equations are detailed in the supplementary material section.

## RESULTS AND DISCUSSION

### Evaluation of the peptide-specific model

The list of specific peptides for 5 body fluids (blood, saliva, semen, urine, and vaginal fluid (VGF)) was constructed with 10 individuals for each fluid with as criteria 80% of sensitivity and 90% of specificity. The resulting list comprises 526 peptides (152 blood peptides, 52 saliva peptides, 208 semen peptides, 100 urine peptides, and 14 vaginal fluid peptides), available in Supplementary Table S1. This list includes peptides belonging to the proteins targeted by the immunochromatographic tests used routinely (hemoglobin, α-amylase, prostate-specific antigen, semenogelin I and II, and Uromodulin) but also many other proteins, which broadens the range of potential targets.[8–11]

After the list was constructed, 10 independent, distinct individuals for each fluid and 78 mixtures were tested (Figure 1, Supplementary Table S4). A test based on the ROC curve (Supplementary Figure S2A) was constructed to maximize sensitivity and specificity. The thresholds on the ratio between the number of peptides found on samples and the number of peptides on the list were established for each fluid: 26.32% for blood, 13.46% for saliva, 32.69% for semen, 3.00% for urine, and 28.57% for vaginal fluid.

The test was evaluated according to several metrics with different significations (see Supplementary Table S4 for details).

Accuracy per class is 100% for blood and saliva and more than 90% for semen, urine and vaginal fluid. Recall (also named sensitivity) is more than 89% for each class, meaning samples typically have the correct label among the predictions. Precision reflects the proportion of correct responses among explicit identifications. Precision is 100% for blood and saliva but slightly lower for semen, urine, and vaginal fluids (98, 88, and 91%), which are sometimes detected even when they are not present. False positives due to limited precision are often associated with mixed samples, which are inherently more difficult to identify correctly than pure samples. For semen, one false positive is linked to a male urine sample that was detected as a mixture of semen and urine. This may be explained by the fact that the individual provided both semen and urine samples; depending on the order of collection, it is plausible that semen was present in the urine sample. The reduced accuracy for saliva is mainly due to a significant number of false positives in mixed samples containing saliva. Urine shows the lowest overall performance, with both a high false positive rate—particularly in samples containing semen or vaginal fluid—and a high false negative rate in mixed samples, leading to the poorest recall. These differences between the fluids can be seen in the detailed results (Supplementary Table S4) and in the heatmap (Figure 1). For vaginal fluid, the challenge lies in the similarity between proteins found in saliva and vaginal fluid, due to their biological nature.[53,54] Consequently, identifying unique peptides specific to vaginal fluid is difficult, contributing to the identification of only 14 specific peptides. This observation was already noted in the study by Zhang et al. (2024), where vaginal fluid and saliva clustered together following hierarchical cluster analysis of proteomic data.[21] Moreover, in the same study, only six tissue-specific proteins were identified for vaginal fluid. In contrast, our analysis revealed peptides corresponding to additional proteins not reported by Zhang et al. (see Supplementary Table S7). For this fluid in particular, a more quantitative approach

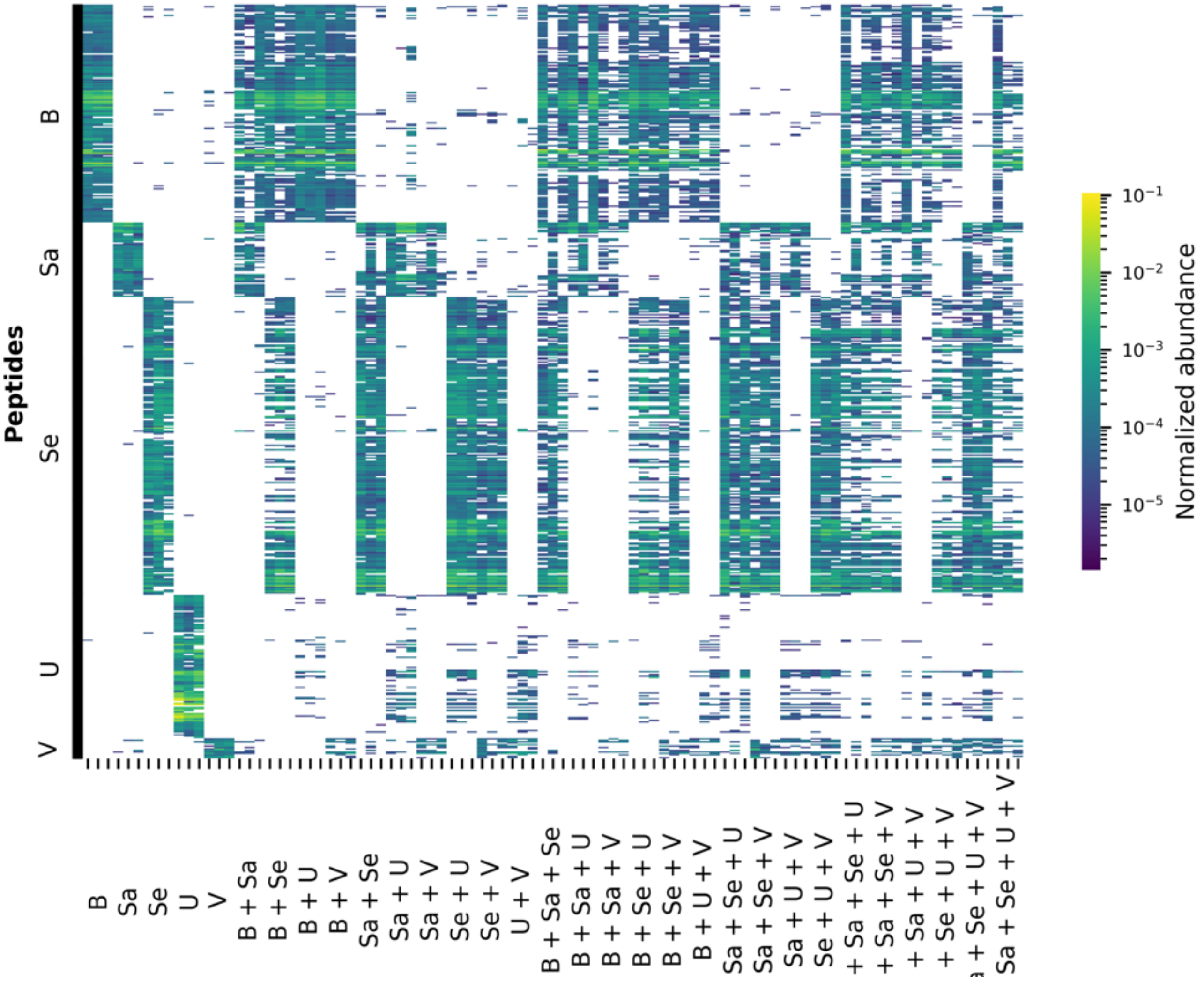

Figure 1: Heatmap representing the normalized intensities of selected peptides for pure body fluids and mixtures. Normalized abundance was obtained by dividing the peptide's abundance by the sum of all the peptides detected in the sample. The peptides selected, shown on the x-axis, are sorted according to the fluid they are specific to and the proteins to which they belong. 3 pure samples from the test set for each fluid identity are presented on the left. Three samples for each mixture are presented on the right. B=Blood, Sa=Saliva, Se=Semen, U=Urine, V=Vaginal fluid

appears necessary, as some proteins are shared with other fluids but present in distinct abundance. As an example, while peptides from digestive proteins like human alpha-amylase are easily identified in saliva due to high concentration, their identification could not be validated in our vaginal fluid samples, unlike in other studies that used larger fluid volumes.[55] Regarding urine, the lower sensitivity observed, particularly in mixtures, may be attributed to its lower protein concentration than other fluids. During sample preparation, mixtures were initially standardized by volume, potentially diluting urine proteins. The relatively low concentration of certain urine peptides, compared to peptides from other fluids, can result in their failure to be detected during our Top 10 DDA bottom-up proteomics analysis. Furthermore, some urine samples tested positive for vaginal fluid, which the anatomical proximity of the female urogenital tract may explain. In addition, one urine sample tested positive for semen, which may be explained by the same hypothesis in men.

With this test, 86.7% of the responses were utterly correct for all samples. However, overall performance is better for pure fluids than for mixed samples.

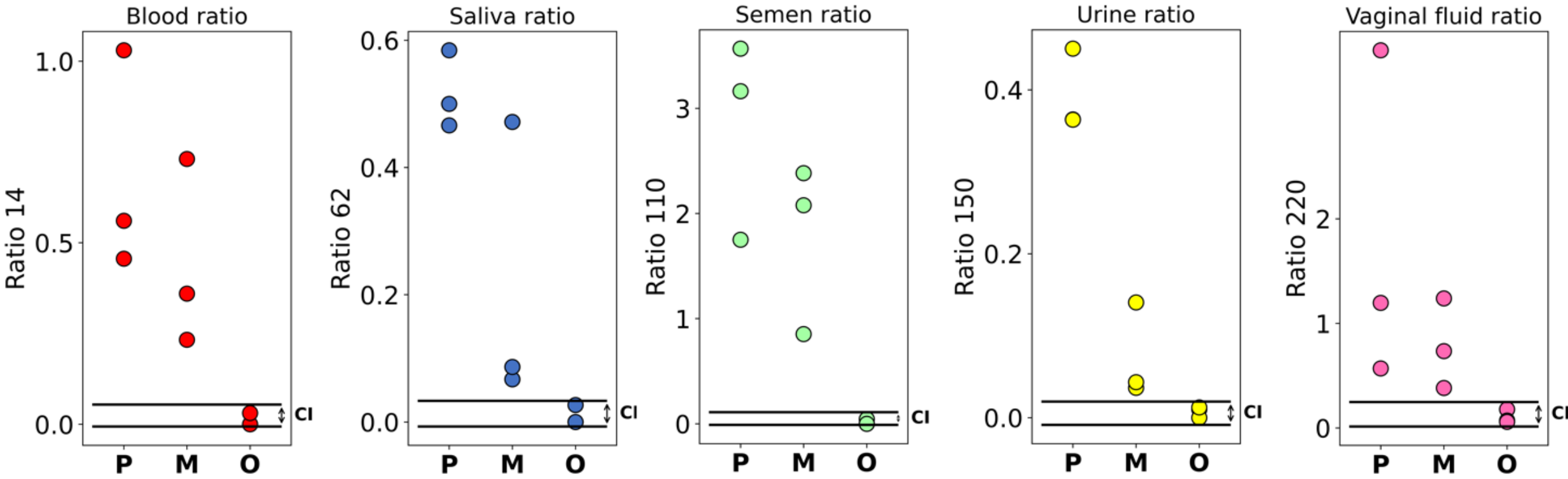

Figure 2: Example ratios per fluid, calculated for selected samples. Each colored circle represents a randomly selected sample. For each fluid, we selected:

- three pure samples containing only that fluid (P),
- three mixed samples containing that fluid along with others (M),
- and three samples (either pure or mixed) that do not contain the fluid (O).

The horizontal lines indicate the 99.99% confidence interval (CI), computed from 10 pure samples of each of the other pure fluids (i.e., excluding the fluid being tested). This CI serves as a reference for the detection test:

- A sample is considered positive for the fluid if its ratio falls outside the CI.

If the ratio falls within the CI, the test is considered negative.

## Evaluation of the quantitative model

A quantitative approach was implemented to detect proteins with fluid-specific abundance variations rather than strict specificity. Direct absolute quantification via mass spectrometry was impractical for complex crime scene samples due to the lack of suitable internal references. Therefore, we used peptide abundance ratios. Applying this method to our initial dataset, we identified 250 significant ratios through ANOVA (Supplementary Table S2). Among the peptides included in the numerator of the ratio, some overlap with tissue-specific peptides, while others do not. These peptides may be present in multiple body fluids, but their relative abundance is characteristic of each fluid.

To test the pertinence of these ratios, 10 new individuals for each fluid and 3 individuals for each possible mixture were used. Figure 2 shows an example of a ratio for each fluid for some samples (fluid of interest pure or in mixture, and other fluids pure or in mixtures). As expected, ratios of mixtures with the selected fluid range between pure fluid of interest and samples without the fluid of interest. A test utilizing the ROC curve (Supplementary Figure S2B) was designed to optimize both sensitivity and specificity. The thresholds for the number of peptides required for positive identification of each fluid were established as follows: 44% for blood, 20% for saliva, 4% for semen, 8% for urine, and 40% for vaginal fluid. Evaluation using the same metrics as the previous test was conducted with the test set (see Supplementary Table S5).

The accuracy of this test was more than 90% for all classes. Performance was slightly lower for vaginal fluid compared to other fluids, notably because vaginal fluid was misidentified in saliva samples, as shown in Supplementary Table S5. As in the previous test, the similarity between vaginal fluid and saliva in terms of protein composition means that the test performs less well for vaginal fluid than for other fluids. Contrary to the previous test, mixtures are better detected than pure fluid, with an accuracy of 75.6% compared to 68.0% for pure fluid

## Evaluation and comparison of machine learning models

### Choice of multilabel classification algorithm

Body fluid identification can be achieved using ML algorithms as a classification task. Compared to classical multiclass classification, identifying body fluid from an unknown trace presents a greater challenge due to the presence of potential fluid mixtures. This complexity makes it a multilabel classification task.

To select the best algorithm for this task, 5 different classifiers (RF, SVM, KNN, LR, and GNB) with 2

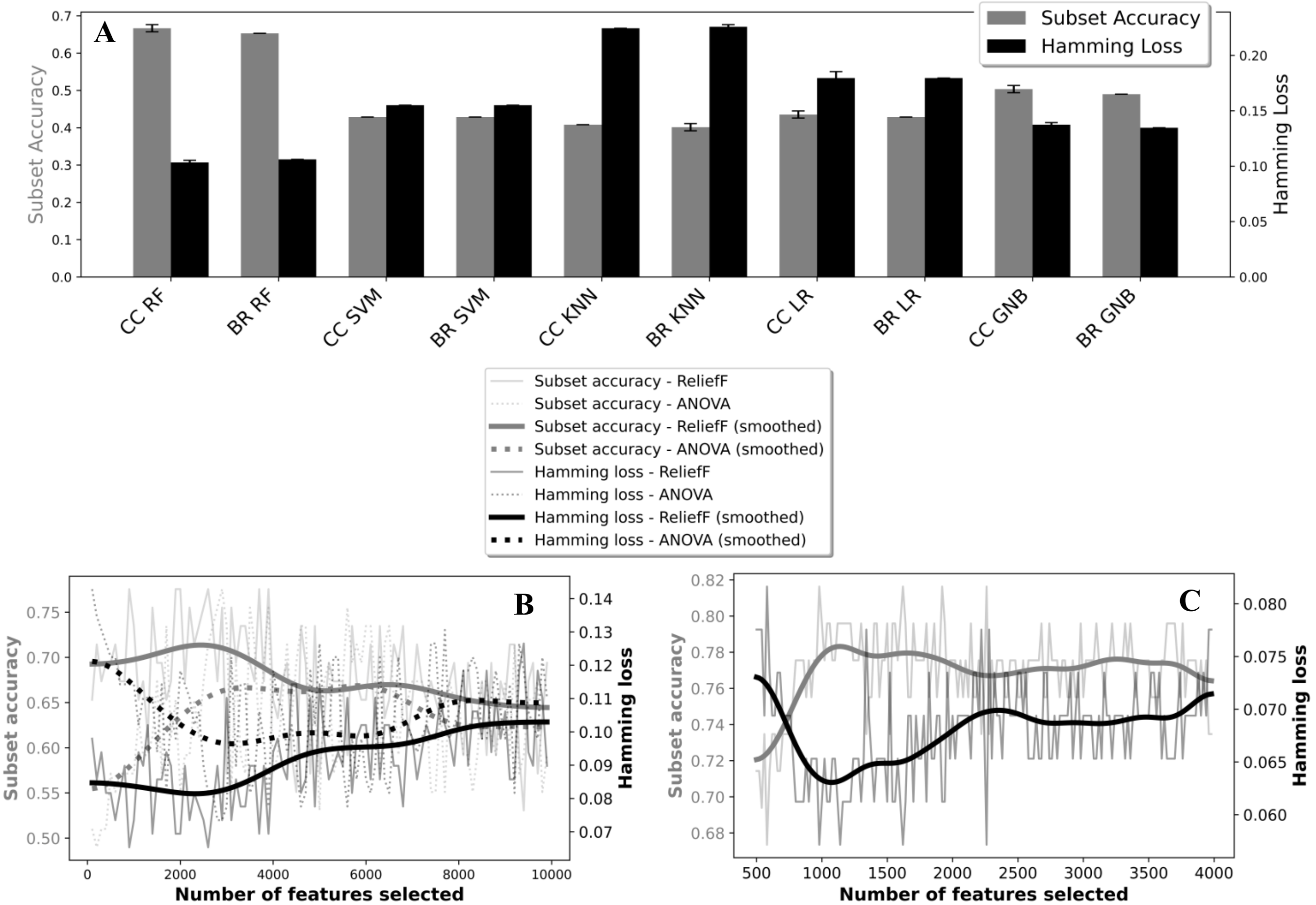

Figure 3: Effect of the choice of the AI algorithm and feature selection on the performance of the model for body fluid recognition.

(A) Performance of the 5 unsupervised classifiers (RF, SVM, KNN, LR, and GNB) on validation data sets after optimization, CC=Classifier Chain, BR=Binary Relevance. Error bars represent the standard deviation on the three replicates of the optimization procedure. (B) Evolution of subset accuracy and Hamming loss of CC-RF for features number from 100 to 10,000 by ReliefF-inspired algorithm or ANOVA test, with a of 100-features steps and Gaussian smoothing fit $\sigma$ = 8. (C) Evolution of subset accuracy and Hamming loss of CC-RF for features number from 500 to 4,000 by ReliefF-inspired algorithm, with 20 feature steps and Gaussian smoothing fit $\sigma$ = 8. Hamming loss and subset accuracy were calculated for each model on validation data after 30 ((A) and (C)) or 5 (B) optimization cycles by the Optuna algorithm (for parameters, see Computation Methods and Programs).

different problem transformation methods (Classifier Chain and Binary Relevance methods for multilabel classification tasks) were optimized and trained on training subsets by minimizing Hamming loss. For multilabel classification tasks with varying misclassification penalties, this approach is appropriate. Misclassifying a urine-blood mixture as urine is a less severe error than misclassifying it as saliva. Therefore, minimizing Hamming loss provides a more comprehensive evaluation of model performance, considering all nuances associated with multilabel predictions. In addition, subset accuracy and Hamming loss were calculated using validation data.

The results presented in Figure 3A show that RF outperformed other algorithms. With RF, the Classifier chain is better than the Binary relevance strategy. Consequently, a Classifier Chain coupled with Random Forest (CC-RF) was used in this study.

Choice of features

By pooling all data from the training set (12 individuals for each pure fluid, 9 negative controls,

and 26 mixtures), 27,664 peptides were detected at least once. However, many peptides were not reproducibly observed and are likely not crucial for body fluid discrimination. Thus, a feature selection filter is necessary to enhance the model classifier. Two different methods were tested on this dataset: a selection based on the p-value obtained from an ANOVA test and the ReliefF algorithm adapted for multilabel feature selection.[36,38] Subset accuracy and Hamming loss for all models after optimization are represented in Figure 3B. Subset accuracy, or "exact match," describes the model's ability to provide exactly the correct explicit prediction; it is a highly stringent metric. Conversely, Hamming loss represents the number of labels that are explicitly incorrect in the test. Consequently, a higher subset accuracy and a lower Hamming loss indicate a better model. The ReliefF-like algorithm outperformed ANOVA-based feature selection, particularly when selecting between 500 and 4,000 features, with higher subset accuracy and lower Hamming loss.

The model's performance increases with the number of features until the number of features becomes too large, which leads to a decrease in performance (Figure 3B). To best observe this phenomenon and choose the optimal number of features, performance was evaluated between 500 and 4,000 features with more cycles (Figure 3C) for the reliefF-based algorithm. After smoothing, maximum subset accuracy is observed at 1,140 features and a minimum Hamming loss at 1,080 features. The ideal number of features was chosen to be equidistant from these two points, which is 1,110 features.

Performance of the final model

Finally, the best model (Classifier chain Random Forest with 1,110 features) was optimized with training and validation sets during 50 cycles of Optuna.[50] The optimization was made by minimizing the Hamming loss on the validation data. This optimization enabled the selection of parameters for Random Forest: 362 estimators, None as max of depth, 8 as the minimum number of samples required to split a node, and 1 as the minimum number of samples needed to be at a leaf node.

The validation and training datasets were compiled, and the selected model was trained with the parameters mentioned earlier. The performance was evaluated on the test dataset with the same metrics as the other models (see Supplementary Table S6).

Although this dataset is slightly smaller than the one used to evaluate the previous models, the proportion of mixtures and pure fluid is similar to the initial test set. The samples used to evaluate this test are available in Supplementary Table S6. Overall, this test outperforms other tests with a subset accuracy of 95.9%. It is noteworthy that there is a subset accuracy of 100% for pure fluid, which is promising for potentially improving mixture results given the small number of mixtures present in the training data (2 biological replications per possible mixture) compared to pure fluids. This result should be interpreted considering the test set includes only 20 pure samples and 26 mixtures compared to the 50 pure samples and 78 mixtures in the initial tests. Still, they are encouraging, given the amount of data available for training.

Comparison of the features used for the 3 different models

Since the models are based on the same data (peptide abundances obtained by mass spectrometry), it is interesting to examine the commonalities among these tests. First, it can be investigated whether some features are shared among some or all of the tests. A Venn diagram is provided in Figure 4A. As shown in Figure 4A, only a few peptides are shared by all algorithms (15 peptides). Additionally, the Random Forest model uses less than a third of the peptides of either the specific peptide model or the ratio model.

To assess the significance of features shared with traditional models in the AI model, Random Forest features can be sorted by their Mean Decrease in Impurity (MDI), which reflects their importance. The histogram depicted in Figure 4B illustrates the distribution of these features based on their weight and the proportion of AI features shared with other models. Generally, there does not seem to be a strong correlation between the importance of features in the AI model and their usage in different models.

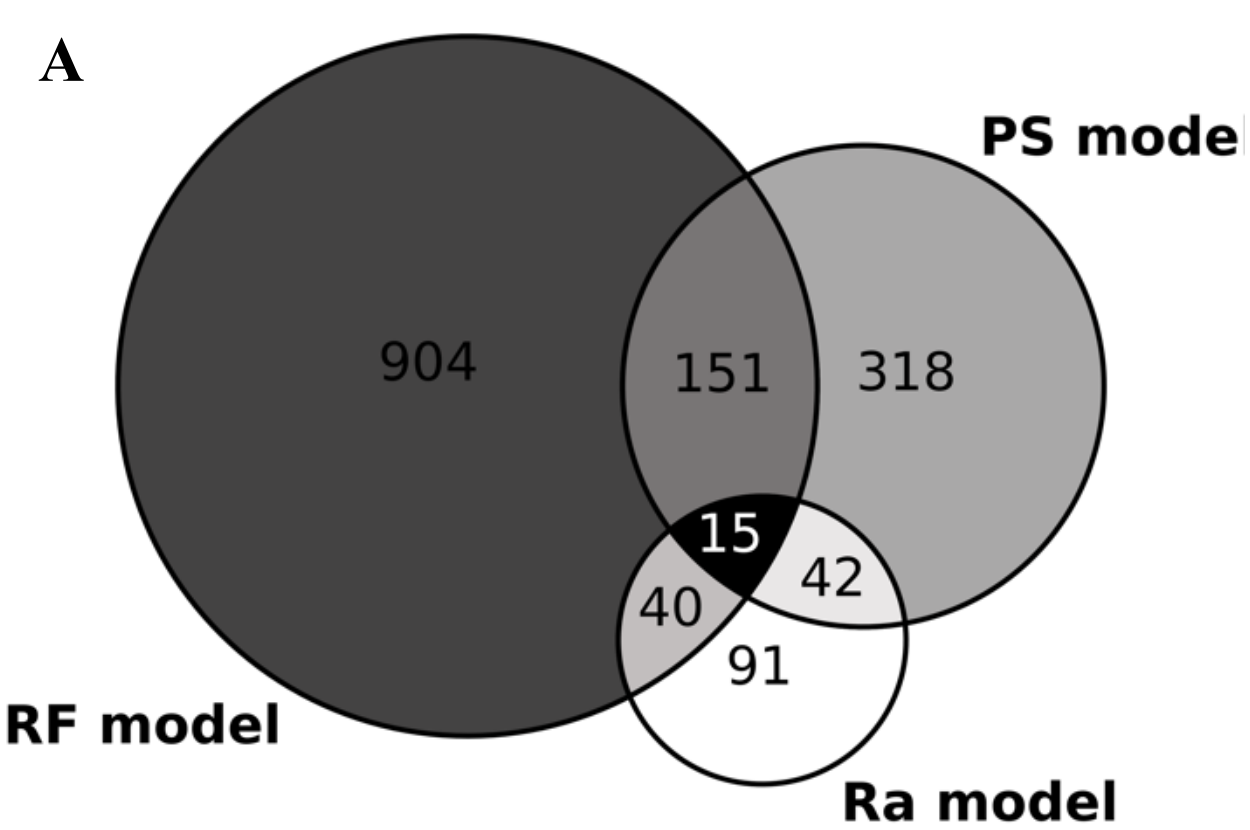

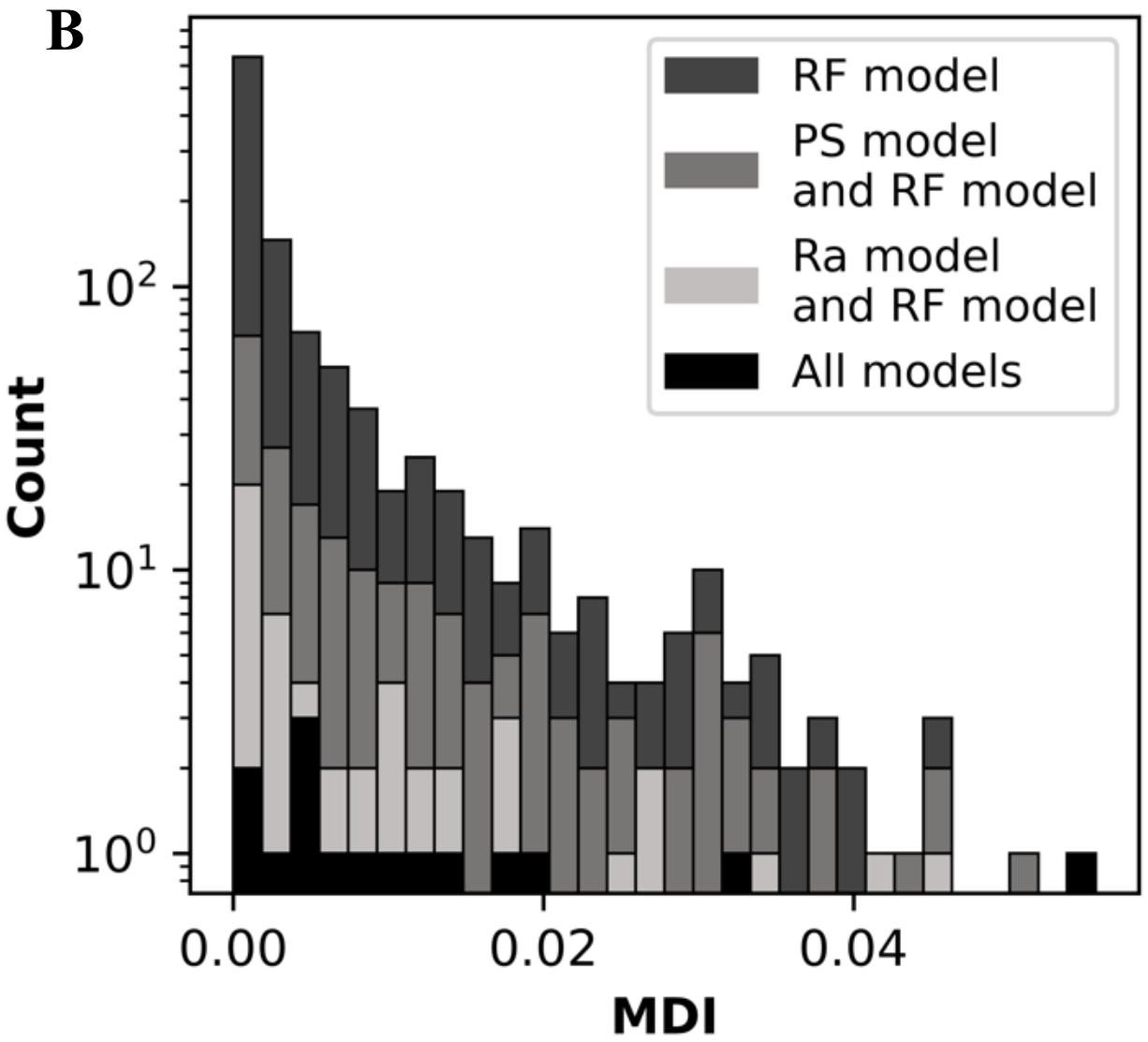

Figure 4: Comparison of features used in each test (PS model = peptide-specific model, Ra model = Ratio-based model, and RF model = Random Forest model)
(A)Venn diagram of the features of the 3 different tests. For the peptide-specific model, features are all the peptides defined in Supplementary Table S1; for the ratio-based model, it is all the peptides implicated in at least one ratio presented in Supplementary Table S2; for the RF model, it is all the 1,110 peptides selected by reliefF-like algorithm detailed Supplementary Table S3. (B) Importance of Random Forest features sorted by MDI (Mean Decrease in Impurity) score

Finally, as the three tests are based on quite different data overall, it can be assumed that their simultaneous use maximizes the chance of results, unlike existing protein tests based on detecting a single target.

## Conclusion

In this study, we developed and evaluated three complementary proteomic strategies for the identification of forensic body fluids using LC-HRMS/MS: peptide-specific biomarkers, peptide abundance ratios, and a machine learning (ML) model. While each approach displayed unique strengths, their combined application represents a significant step forward, particularly for the challenging problem of discriminating body fluids in mixtures.

The specific peptide-based approach provides a robust and transferable list of diagnostic peptides that can be readily applied even on less sensitive instruments. Importantly, the method does not rely on the detection of a single protein such as α-amylase, but on a panel of peptides whose combined presence increases specificity and reduces the risk of false positives. The ML model was built on a Random Forest classifier, achieved 100% accuracy on pure fluids and showed promising performance on mixtures, underscoring the potential of AI-driven proteomics. The abundance ratio method was less robust, but, when used in combination with biomarkers or ML, it may still provide complementary insights.

In forensics, the application of these approaches to mixtures is particularly promising. Our results show that both the biomarker panel and ML can successfully detect the presence of multiple fluids, though interpretive caution is required. As in other forensic assays, the distinction between presumptive and confirmatory conclusions remains critical. Instead of reporting the presence of a single protein marker (e.g., α-amylase, mucins, PSA), we interpret peptide patterns. For example, we prefer to state 'saliva cannot be confirmed' rather than 'saliva confirmed' when only limited or overlapping markers are detected, to ensure that interpretations remain scientifically defensible in court.

We also recognize important limitations of our study. Even though we attempted to mimic casework conditions using swabs, our experiments were performed on neat body fluids and laboratory-prepared mixtures, which do not fully capture the complexity of forensic samples. Real-life samples can degrade due to environmental factors, substrate effects, and biological variability, which can compromise detection. Moreover, while statistically robust, the ML model was trained on samples prepared in the laboratory. These results may not extrapolate easily or successfully to casework without retraining. To overcome these limitations, future work should include real samples of known composition or mock stains prepared under conditions that better mimic forensic evidence, such as degraded materials or mixtures of varying ratios. This is a long-term effort which is already underway.

Ultimately, integrating proteomic workflows with other forensic tools (e.g., DNA analysis), extending data processing to other taxonomies (such as microbiota and other species, including rodents or animals found in the area of interest), and developing user-friendly software for interpretation will be key to enabling routine adoption. Taken together, our findings highlight the power of combining targeted peptide biomarkers with ML-based pattern recognition. With further validation under case-like conditions, this dual strategy has the potential to provide forensic laboratories with a reliable, interpretable, and comprehensive approach to body fluid identification in both single and mixed samples.

## ASSOCIATED CONTENT

### Raw and metadata

The mass spectrometry proteomics data have been deposited to the ProteomeXchange Consortium via the PRIDE partner repository with the dataset identifier PXD053891.[35]

### Code availability

Scripts, sample descriptions, and processed data used for analyses carried out in this study have been deposited on the GitHub repository at https://github.com/SMBP-lab/Body-fluid-identification-proteomics.

### Supporting Information

The Supporting Information is available in the supplementary data file.

- Supplementary materials, sex distribution, and ROC curves (PDF)
- Peptides used for each test and performance of each test (XLSX)

## AUTHOR INFORMATION

### Corresponding Author

* Joelle Vinh: Spectrométrie de Masse Biologique et Protéomique, ESPCI-PSL, CNRS, UAR 2051, 10 Rue Vauquelin, 75005, Paris, France
joelle.vinh@espci.fr

### Author Contributions

All authors have approved the final version of the manuscript.

## ACKNOWLEDGMENT

This work was supported by funds from the Forensic Science Laboratory of the French Gendarmerie (IRCGN: Institut de Recherche de la Gendarmerie Nationale). Conseil Regional d'Île-de-France subsidized mass spectrometry equipment (Sesame 2010 10022268, Sesame 2018 EX039194).

TOC graphic

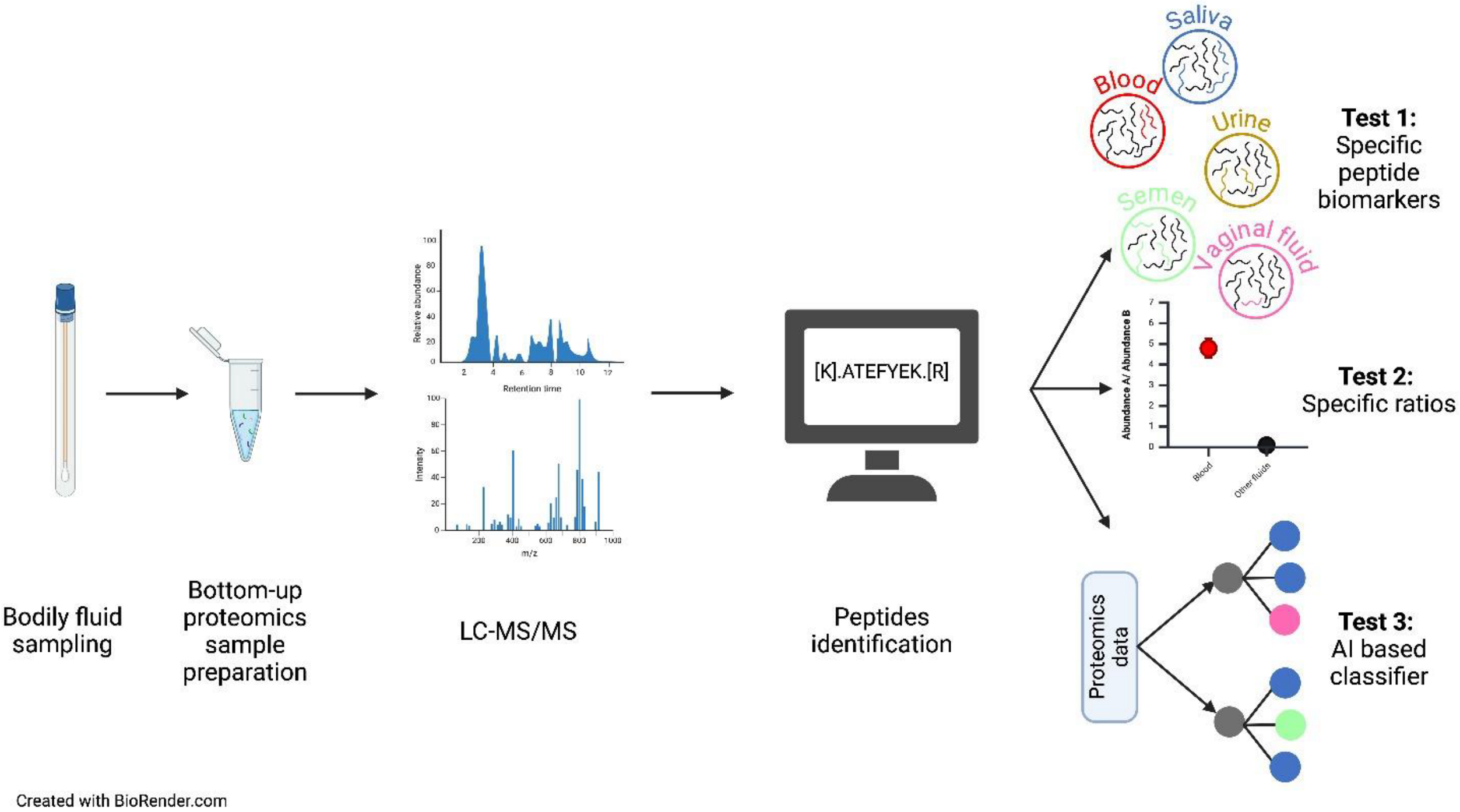

Saliva
Blood
Urine
Semen
Vaginal fluid
Test 1:
Specific peptide biomarkers
Test 2:
Specific ratios
Abundance A/ Abundance B
Blood
Other fluids
Test 3:
AI based classifier
Proteomics data
Relative abundance
Retention time
Intensity
m/z
[K].ATEFYEK.[R]
Bodily fluid sampling
Bottom-up proteomics sample preparation
LC-MS/MS
Peptides identification
Created with BioRender.com

# The Identification of Biological Stains at Crime Scenes: A Promising Role for Proteomics and Machine Learning. Associated Content

Anna Rosenberg[†‡], Stéphanie Laurent[¶†], Esther Morandeau[¶†], Alix Munoz[†], and Joelle Vinh[*†]

[†]Spectrométrie de Masse Biologique et Protéomique, ESPCI-PSL, CNRS, UAR 2051, Rue Vauquelin, 75005, Paris, France

[‡]Department of biology, Ecole Normale Supérieure Paris-Saclay, Avenue des Sciences, 91190, Gif-sur-Yvette, France

[¶]Département Environnement Incendies Explosifs, Institut de Recherche Criminelle de la Gendarmerie Nationale, Boulevard de l'Hautil, 95000, Cergy-Pontoise, France

.

## SUPPORTING INFORMATION

### METRICS USED FOR MODEL EVALUATION

Accuracy, F1-score, precision, and recall were calculated for each fluid as in equations (1), (2), (3) and (4):

For fluid j (j∈ [[1; 5]])

$$Accuracy_j = \frac{TP_j+TN_j}{TP_j+TN_j+FP_j+FN_j} \quad (1)$$

$$Precision_j = \frac{TP_j}{TP_j+FP_j} \quad (2)$$

$$Recall_j = \frac{TP_j}{TP_j+FN_j} \quad (3)$$

$$F1_j = 2\frac{Precision_j x Recall_j}{Preision_j+Recall_j} \quad (4)$$

where :

the number of True Positives for class j: TPj

the number of True Negatives for class j: TNj

the number of False Positives for class j: FPj

the number of False Negatives for class j: FNj

Micro, macro, and weighted F1-score, precision, and recall were calculated according to equations (5), (6) and (7):

$$B_{macro} = \frac{1}{q}\sum_{j=1}^{q} B(TP_j, FP_j, TN_j, FN_j) \quad (5)$$

$$B_{micro} = B\left[\sum_{j=1}^{q} (TP_j), \sum_{j=1}^{q} (FP_j), \sum_{j=1}^{q} (TN_j), \sum_{j=1}^{q} (FN_j)\right] \quad (6)$$

$$B_{weighted} = \sum_{j=1}^{q} w_j \times B(TP_j, FP_j, TN_j, FN_j) \quad (7)$$

where :

$B(TP_j, FP_j, TN_j, FN_j)$ rerepresent some specific binary classification metric (B ∈ Accuracy, Precision, Recall, F1-score)

q = number of class labels = 5 (blood, saliva, semen, urine, and vaginal fluid)

$w_j = \frac{N_j}{N} =$ weight of class j

Nj = number of samples with class j label

N = Total number of samples

Subset accuracy was also calculated as the percentage of samples with all labels classified correctly.23 Subset accuracy is defined in equation (8):

$$subsetacc(h) = \frac{1}{p}\sum_{i=1}^{p} \quad h(xi) = Y_i \quad (8)$$

where :

p = number of test samples

h(xi)=prediction made by the model for sample xi

Yi = real class of sample xi

Subset accuracy, also known as "exact match," represents the model's ability to provide exactly the correct prediction; it is an extremely stringent metric. Models were also evaluated with Hamming loss score defined in equation (9):

$$h\ hloss(h) = \frac{1}{p}\sum_{i=1}^{p} \quad \frac{1}{q}|h(xi)\Delta Yi| \quad (9)$$

where :

p = number of test samples

q = number of class labels = 5 (blood, saliva, semen, urine, and vaginal fluid

$h(x_i)$=prediction made by the model for sample $x_i$

$Y_i$ = real class of sample $x_i$

$|h(x_i) \Delta Y_i|$ = symmetric difference between the two sets. For example, if the prediction is AB and the actual class is C, it will be equal to 3

Model performance improves as the Hamming loss decreases.

SUPPLEMENTARY FIGURES

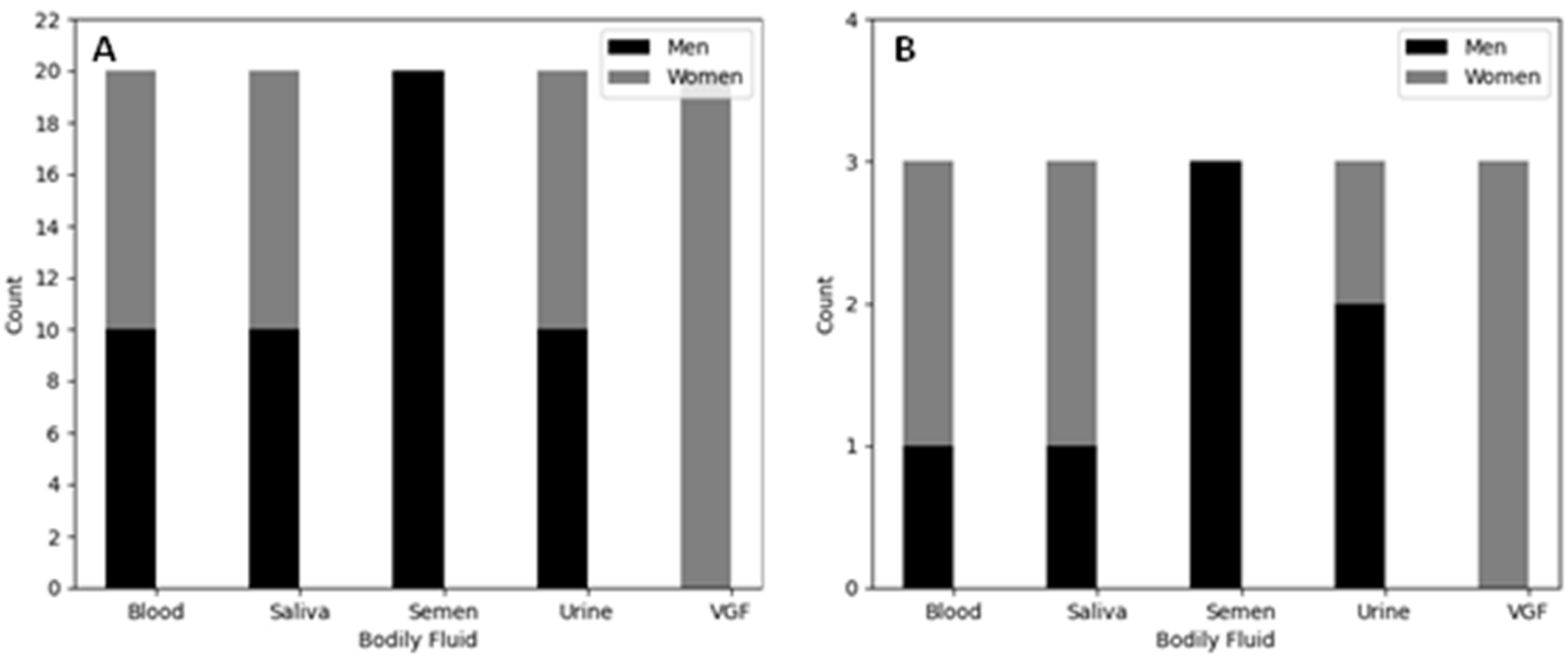

Supplementary Figure S1: Number of participants and sex distribution of cohort (A) Number of participants and sex distribution of cohort used for pure bodily fluids; (B)Number of participants and sex distribution used for mixed samples

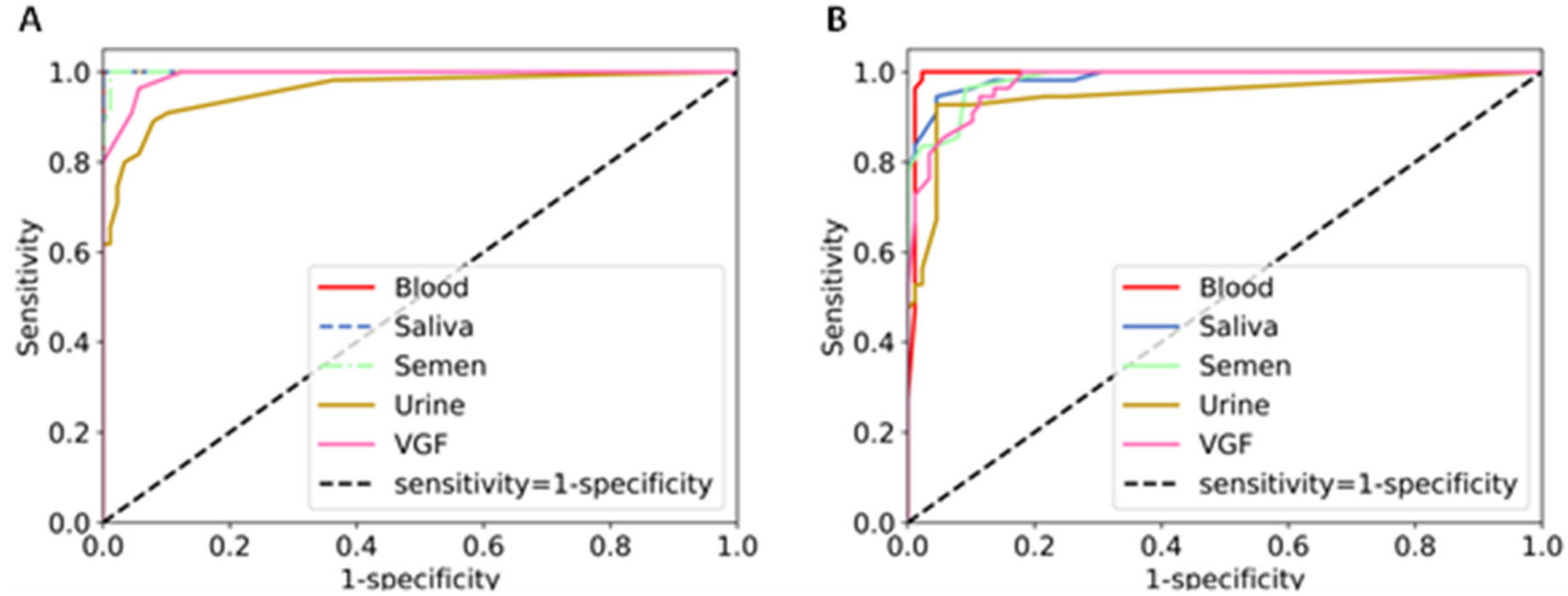

Supplementary Figure S2: ROC curve of (A) the peptide list-based test for all fluids; (B) the peptide ratio-based test for all fluids. This curve was constructed from test data after peptide selection in A, ratio selection in B. The top left corner corresponds to maximum sensitivity and specificity, where $Sensitivity_j = Recall_j = \frac{TP_j}{TP_j + FN_j}$ and $Specificity_j = \frac{TN_j}{TN_j + FP_j}$.

.

**SUPPLEMENTARY TABLES.**

In Supplementary Tables S1, S2 and S3, peptide cells are color coded with Red=Blood, Blue=Saliva, Green=Semen, Yellow=Urine and Pink=Vaginal Fluid

**Supplementary Table S1: Peptides selected for list-based test**

Peptides selected were obtained from 10 independent individuals for each fluid with criteria of 80% of sensitivity and 90% of specificity.

**Supplementary Table S2: Ratios selected for quantitative test**

The selected ratios were obtained from 10 independent individuals for each fluid, and 50 ratios per fluid were selected depending on the p-value result of the ANOVA test.

**Supplementary Table S3: Peptides selected by reliefF-like method for AI-based test**

Peptides selected were obtained from the training set by using a reliefF-like algorithm for multi-label classification.

**Supplementary Table S4: Detailed results of performance test for list-based test**

**Supplementary Table S5: Detailed results of performance test for quantitative test**

**Supplementary Table S6: Detailed results of performance test for AI-based test**

**Supplementary Table S7: Comparison of proteins found in the article by Zhang et al (2024) with proteins associated with our target peptides**

## AUTHOR INFORMATION

### Corresponding Author

* Joelle Vinh: Spectrométrie de Masse Biologique et Protéomique, ESPCI-PSL, CNRS, UAR 2051, 10 Rue Vauquelin, 75005, Paris, France
joelle.vinh@espci.fr

### Author Contributions

All authors have approved the final version of the manuscript.

## ACKNOWLEDGMENT

This work was supported by funds from the Forensic Science Laboratory of the French Gendarmerie (IRCGN: Institut de Recherche de la Gendarmerie Nationale). Conseil Regional d'Île-de-France subsidized mass spectrometry equipment (Sesame 2010 10022268, Sesame 2018 EX039194).